\documentclass[runningheads]{llncs}
\usepackage[T1]{fontenc}
\usepackage{graphicx}
\usepackage{amssymb}
\usepackage{booktabs}
\usepackage{array}
\usepackage{multirow}
\usepackage[colorlinks,linkcolor=blue,citecolor=blue,urlcolor=blue]{hyperref}
\usepackage{color}
\begin{document}
\title{Lesion-aware Dynamic Kernel for Polyp Segmentation}

\author{Ruifei Zhang \inst{1} \and 
Peiwen Lai\inst{1} \and
Xiang Wan\inst{2,4} \and
De-Jun Fan\inst{3} \and
Feng Gao\inst{3} \and\\
Xiao-Jian Wu\inst{3} \and
Guanbin Li\inst{1,2}\thanks{Corresponding author is Guanbin Li.}
}


%
\authorrunning{R. Zhang et al.}
%

\institute{
 $^1$ School of Computer Science and Engineering, Sun Yat-sen University, \\Guangzhou, China\\
 \email{liguanbin@mail.sysu.edu.cn}\\
 $^2$ Shenzhen Research Institute of Big Data, Shenzhen, China\\
 $^3$ The Sixth Affiliated Hospital, Sun Yat-sen University, Guangzhou, China\\
 $^4$ Pazhou Lab, Guangzhou, China}

%
%
%
%
%
\maketitle              
\begin{abstract}
Automatic and accurate polyp segmentation plays an essential role in early colorectal cancer diagnosis. However, it has always been a challenging task due to 1)~the diverse shape, size, brightness and other appearance characteristics of polyps, 2)~the tiny contrast between concealed polyps and their surrounding regions. To address these problems, we propose a lesion-aware dynamic network (LDNet) for polyp segmentation, which is a traditional u-shape encoder-decoder structure incorporated with a dynamic kernel generation and updating scheme. Specifically, the designed segmentation head is conditioned on the global context features of the input image and iteratively updated by the extracted lesion features according to polyp segmentation predictions. This simple but effective scheme endows our model with powerful segmentation performance and generalization capability. Besides, we utilize the extracted lesion representation to enhance the feature contrast between the polyp and background regions by a tailored lesion-aware cross-attention module (LCA), and design an efficient self-attention module (ESA) to capture long-range context relations, further improving the segmentation accuracy. Extensive experiments on four public polyp benchmarks and our collected large-scale polyp dataset demonstrate the superior performance of our method compared with other state-of-the-art approaches. The source code is available at \url{https://github.com/ReaFly/LDNet}.

\end{abstract}
\section{Introduction}
Colorectal Cancer (CRC) is one of the most common cancer diseases around the world~\cite{Siegel2022cancer}. However, actually, most CRC starts from a benign polyp and gets progressively worse over several years. Thus, early polyp detection and removal make essential roles to reduce the incidence of CRC. In clinical practice, colonoscopy is a common examination tool for early polyp screening. An accurate and automatic polyp segmentation algorithm based on colonoscopy images can greatly support clinicians and alleviate the reliance on expensive labor, which is of great clinical significance.

However, accurate polyp segmentation still remains challenge due to the diverse but concealed characteristics of polyps. Early traditional approaches~\cite{mamonov2014automated,tajbakhsh2015automated} utilize hand-craft features to detect polyps, failing to cope with complex scenario and suffering from high misdiagnosis rate. With the advance of deep learning technology, plenty of CNN-based methods are developed and applied for polyp segmentation. Fully convolution network~\cite{long2015fully} is first proposed for semantic segmentation, and then its variants~\cite{brandao2017fully,akbari2018polyp} also make a great breakthrough in the polyp segmentation task. UNet~\cite{ronneberger2015u} adopts an encoder-decoder structure and introduces skip-connections to bridge each stage between encoder and decoder, supplying multi-level information to obtain a high-resolution segmentation map through successive up-sampling operations. UNet++~\cite{zhou2018unet++} introduces more dense and nested connections, aiming to alleviate the semantic difference of features maps between encoder and decoder.
Recently, to better overcome the above mentioned challenges, some networks specially designed for polyp segmentation task have been proposed. For example, PraNet~\cite{fan2020pranet} adopts a reverse attention mechanism to mine finer boundary cues based on the initial segmentation map. ACSNet~\cite{zhang2020adaptive} adaptively selects and integrates both global contexts and local information, achieving more robust polyp segmentation performance. CCBANet~\cite{nguyen2021ccbanet} proposes the cascading context and the attention balance modules to aggregate better feature representation. SANet~\cite{wei2021shallow} designs the color exchange operation to alleviate the color diversity of polyps, and proposes a shallow attention module to select more useful shallow features, obtaining comparable segmentation results. 
However, existing methods mainly focus on enhancing the network's lesion representation from the view of feature selection~\cite{nguyen2021ccbanet,zhang2020adaptive,fan2020pranet} or data augmentation~\cite{wei2021shallow}, and no attempts have been made to consider the structural design of the network from the perspective of improving the flexibility and adaptability of model feature learning, which limits their generalization. 

To this end, we design a Lesion-aware Dynamic Network (LDNet) for the polyp segmentation task. Inspired by~\cite{he2019dynamic,zhang2021k}, we believe that a dynamic kernel can adaptively adjust parameters according to the input image, and thus achieving stronger feature exploration capabilities in exchange for better segmentation performance. Specifically, our unique kernel (also known as segmentation head) is dynamically generated basing on the global features of the input image, and generates one polyp segmentation prediction in each decoder stage. Accordingly, these segmentation results serve as clues to extract refined polyp features, which in turn update our kernel parameters with better lesion perception. For some complex polyp regions, the dynamic kernel generation and update mechanism we designed can step-wisely learn and mine discriminative regional features and gradually improve the segmentation results, enhancing the generalization of the model. 
Besides, we design two attention modules, \emph{i.e.,} Efficient Self-Attention (ESA) and Lesion-aware Cross-Attention (LCA). The former is used to capture global feature relations, while the latter is designed to enhance feature contrast between lesions and other background regions, further improving the segmentation performance. In summary, the contributions of this paper mainly include three folds: (1)~We design a lesion-aware dynamic network for polyp segmentation. The introduction of a dynamic kernel generation and update mechanism endows the model with generalizability to discriminate polyp regions with diverse shapes, sizes, and appearances. 
(2)~Our tailored ESA and LCA modules enhance the polyp feature representation, which helps to mine concealed polyps with low visual contrast. (3)~Extensive experiments on four public polyp benchmarks and our collected large-scale polyp dataset demonstrate the effectiveness of our proposed method.

\begin{figure}[tp]
\centering
\includegraphics[width=0.9\textwidth]{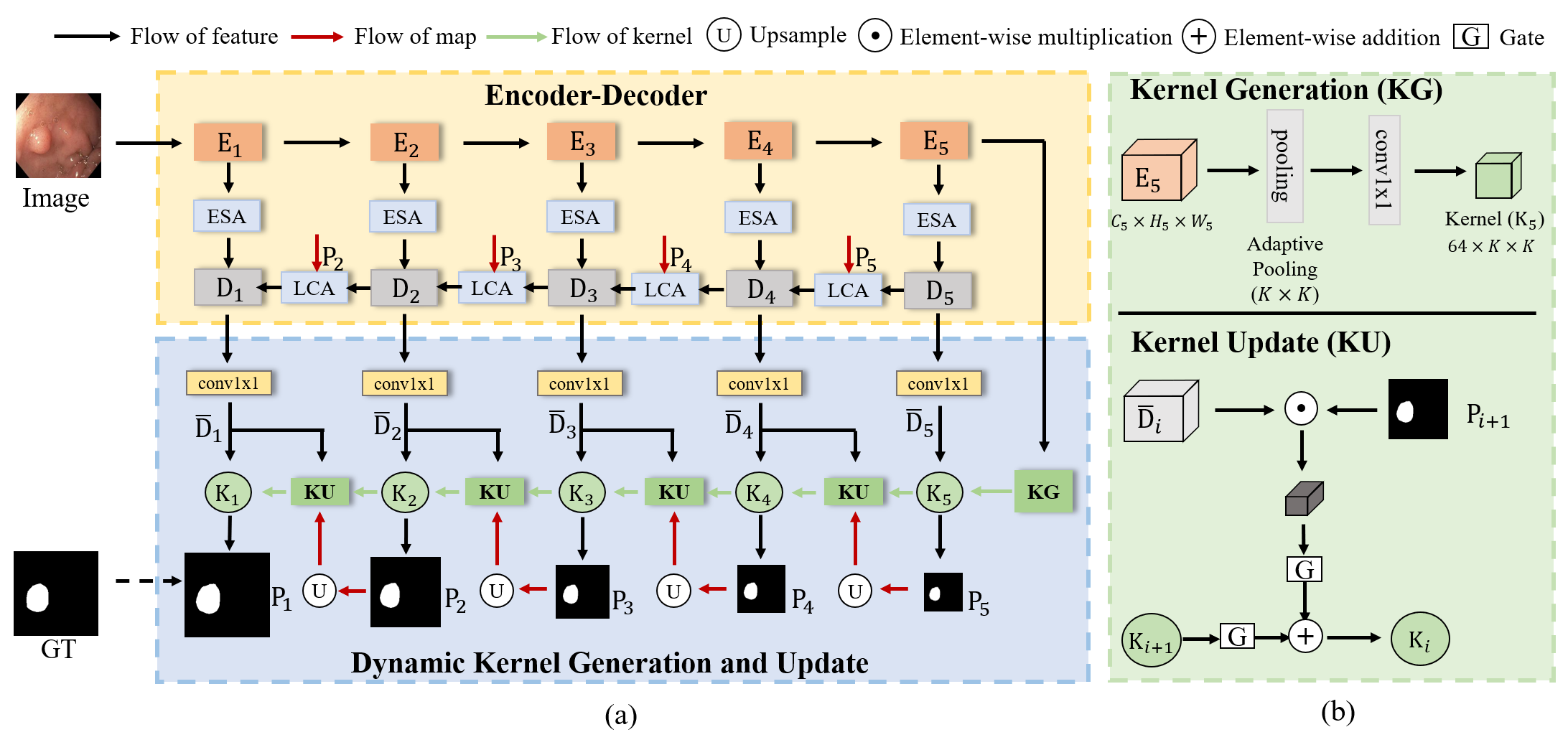} 
\caption{(a) Overview of our LDNet. (b) Illustration of kernel generation and update.}
\label{fig1}
\end{figure}

\section{Methodology}
The overview of our LDNet is shown in Fig.~\ref{fig1}, which is a general encoder-decoder structure, incorporated with our designed dynamic kernel scheme and attention modules. The Res2Net~\cite{gao2019res2net} is utilized as our encoder, consisting of five blocks. The generated feature map of each block is denoted as $\{\mathbf E_i\}_{i=1}^5$. Accordingly, five-layer decoder blocks are adopted and their respective generated features are defined as~$\{\mathbf D_i\}_{i=1}^5$. $1\times1$ convolution is utilized to unify the dimension of $\mathbf D_i$ to 64, denoted as ${\mathbf {\bar D}}_i$, which are adaptive to subsequent kernel update operations. 
In contrast to previous methods~\cite{zhang2020adaptive,fang2019selective,wei2021shallow,fan2020pranet} with a static segmentation head, which is agnostic to the input images and remains fixed during the inference stage, we design a dynamic kernel as our segmentation head. The dynamic kernel is essentially a convolution operator used to produce segmentation result, but its parameters are initially generated by the global feature ${\mathbf E}_5$ of the input, and iteratively updated in the multi-stage decoder process based on the current decoder features ${\mathbf {\bar D}}_{i}$ and its previous segmentation result $\mathbf P_{i+1}$, which is employed to make a new prediction $\mathbf P_{i}$. For the convenience of expression, we denote the sequential updated kernels as $\{\mathbf K_i\}_{i=1}^5$. Each segmentation prediction is supervised by the corresponding down-sampled Ground Truth, and the prediction $\mathbf P_1$ of the last decoder stage is the final result of our model. We detail the dynamic kernel scheme and attention modules in the following sections.
\subsection{Lesion-aware Dynamic Kernel}
\subsubsection{Kernel Generation}
Dynamic kernels can be generated in a variety of ways and have been successfully applied in many fields~\cite{zhang2021k,he2019dynamic,zhang2021dodnet,pang2020hierarchical}. In this paper, We adopt a simple but effective method to generate our initial kernel. As shown in Fig.1, given the global context feature ${\mathbf E}_5$, we first utilize an adaptive average pooling operation to aggregate features into a size of $K \times K$, and then perform one $1\times1$ convolution to produce the initial segmentation kernel with a reduced dimension of $64$. To be consistent with the sequence of decoder, we denote our initial kernel as $\mathbf K_5 \in \mathbb{R}^{1\times 64 \times K \times K}$. $\mathbf K_5$ is acted on the unified decoder features ${\mathbf {\bar D}}_5$ to generate the initial polyp prediction $\mathbf P_5$.
\subsubsection{Kernel Update}
Inspired by~\cite{zhang2021k}, we design an iterative update scheme based on the encoder-decoder architecture to improve our dynamic kernel. Given the $i$-th unified decoder features ${\mathbf {\bar D}}_i \in \mathbb{R}^{64 \times H_i \times W_i}$ and previous polyp segmentation result $\mathbf P_{i+1} \in \mathbb{R}^{1 \times H_{i+1} \times W_{i+1}}$, we first extract lesion features as:
\begin{equation}
\mathbf F_i = \sum^{H_i}\sum^{W_i}{up}_2(\mathbf P_{i+1}) \circ {\mathbf {\bar D}}_i,
\label{equ1}
\end{equation}
where $up_2$ denotes up-sampling the prediction map by a factor of 2 to keep a same size with feature map. `$\circ$' represents the element-wise multiplication with broadcasting mechanism.

The essential operation of the kernel update is to integrate the lesion representations extracted by the current decoder features into previous kernel parameters. In this way, the kernel can not only perceive the lesion characteristics to be segmented in advance, but gradually incorporate multi-scale lesion information, thus enhancing its discrimination ability for polyps. Since the previous polyp prediction may be inaccurate, as in~\cite{zhang2021k}, we further utilize a gate mechanism to filter the noise in lesion features and achieve an adaptive kernel update. The formulation is:
\begin{equation}
\mathbf K_i = \mathbf G_i^F \circ \phi_1(\mathbf F_i) + \mathbf G_i^K \circ \phi_2(\mathbf K_{i+1}) ,
\label{equ2}
\end{equation}
where $\phi_1$ and $\phi_2$ denote linear transformations. $\mathbf G_i^F$ and $\mathbf G_i^K$ are two gates, which are obtained by the element-wise multiplication between the variants of $\mathbf F_i$ and $\mathbf K_{i+1}$ followed by different linear transformation and Sigmoid function ($\sigma$), respectively:
\begin{equation}
\mathbf G_i = \phi_3(\mathbf F_i) \circ \phi_4(\mathbf K_{i+1}) 
\label{equ3}
\end{equation}
\begin{equation}
\mathbf G_i^K = \sigma(\phi_5(\mathbf G_i)), 
\mathbf G_i^F = \sigma(\phi_6(\mathbf G_i))
\label{equ4}
\end{equation}

The updated kernel $\mathbf K_i$ is acted on the specific decoder feature to make a new prediction $\mathbf P_i$. Both of them are sent to the ($i-1$)-th decoder stage to iteratively perform the above update scheme.
\begin{figure}[htp]
\centering
\includegraphics[width=0.67\textwidth]{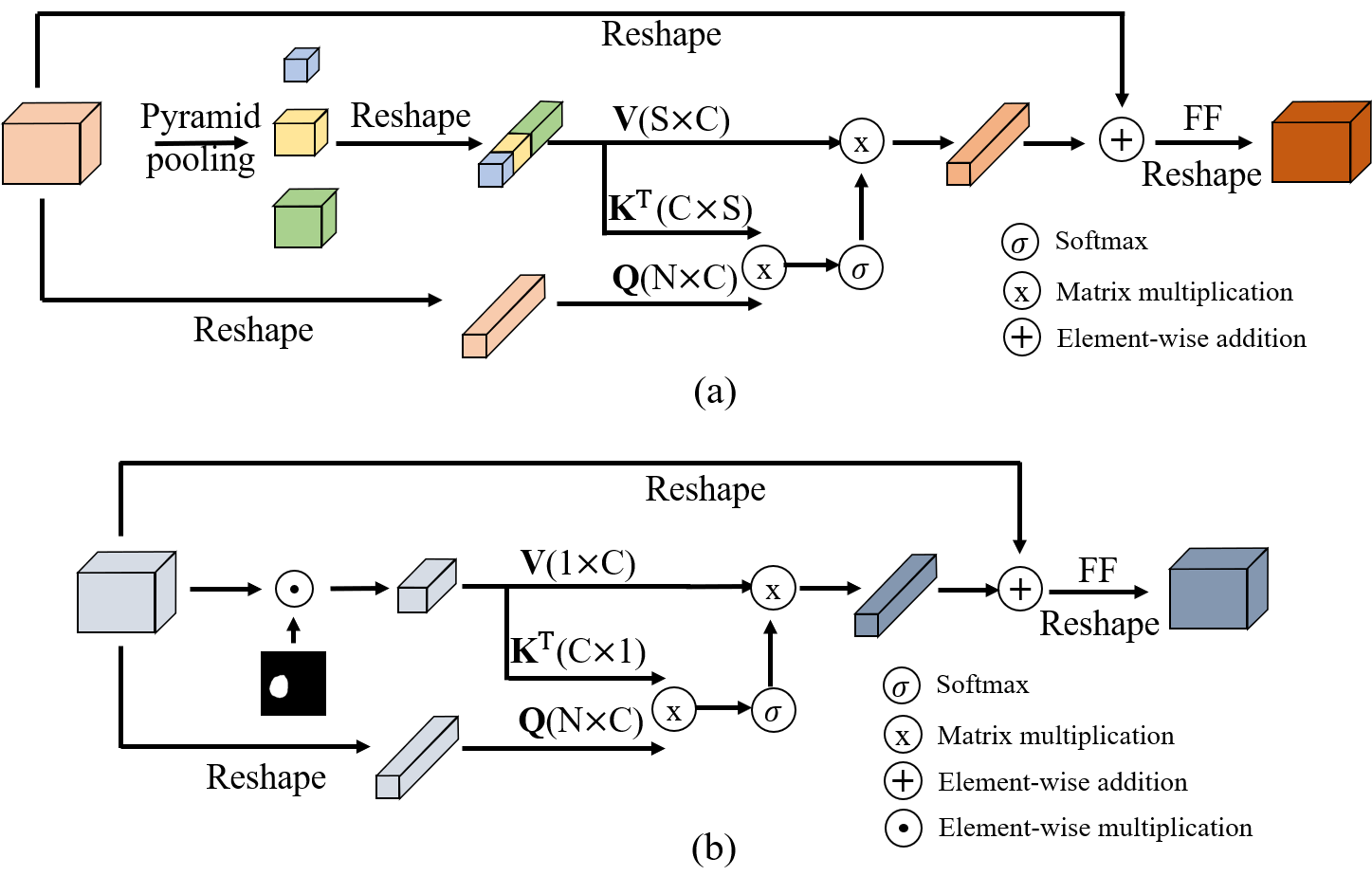} 
\caption{(a) Illustration of ESA. (b) Illustration of LCA. FF denotes the feed-forward layer. We omit the residual addition between the input and output of FF for simplicity.}
\label{fig2}
\end{figure}
\subsection{Attention Modules}
\subsubsection{Efficient Self-Attention}
Self-attention mechanism is first proposed in Transformer~\cite{vaswani2017attention}, and recently has played a significant role in many tasks~\cite{dosovitskiy2020image,carion2020end} due to its strong long-range modeling capability, however is criticized for prohibitive computation and memory cost. To overcome these challenges, we borrow the idea from~\cite{zhao2017pyramid,zhu2019asymmetric} and design our ESA module. As shown in Fig.~\ref{fig2}, we follow the component of Transformer but replace the original self-attention with our ESA layer, followed by a feed-forward layer and a reshaping operation. We also perform a multi-head parallel scheme to further improve the performance. Specifically, given one encoder feature map $\mathbf E_i \in \mathbb{R}^{C_i\times H_i \times W_i}$, details of our ESA layer are formulated as follows:
\begin{equation}
{\rm ESA}(\mathbf E_i) = \phi_o({\rm concat} ({\rm head}^0,...,{\rm head}^n)),
\label{equ5}
\end{equation}
\begin{equation}
{\rm head}^{j} = {\rm Attention}(\phi^j_q(\mathbf Q), \phi^j_k(\mathbf K), \phi^j_v(\mathbf V)),
\label{equ6}
\end{equation}
where $\phi_o$, $\phi^j_q$, $\phi^j_k$, $\phi^j_v$ denote the linear projections, and $n$ is the number of heads. $\mathbf Q \in \mathbb{R}^{N_i\times C_i} (N_i=H_i\times W_i)$ is reshaped from the $\mathbf E_i$. $\mathbf K$, $\mathbf V \in \mathbb{R}^{S\times C_i}$ are obtained by the pyramid pooling operation~\cite{zhao2017pyramid}, which includes $1\times 1$, $3\times 3$, $5\times 5$ adaptive average pooling to down-sample the feature map, followed by reshaping and concatenating operations. Thanks to such a sampling process, we utilize fewer representative global features to perform the standard attention~\cite{vaswani2017attention}, not only introducing global relations to original features, but significantly saving the computation overhead ($S=1\times 1 + 3\times 3 + 5\times 5 \ll N_i$). Attention($\cdot$) is formulated as:
\begin{equation}
{\rm Attention}(\mathbf q,\mathbf k,\mathbf v)={\rm softmax}(\frac{\mathbf q\mathbf k^\mathrm{T}}{\sqrt{d_k}})\mathbf v,
\label{equ7}
\end{equation}
where $d_k$ is the dimension of each head, equal to $\frac{C_i}{n}$.
\subsubsection{Lesion-aware Cross-Attention}
Besides our lesion-aware dynamic kernel, the predicted polyp result is also utilized to enhance the features. Specifically, given the decoder feature $\mathbf D_i \in \mathbb{R}^{C_i\times H_i \times W_i}$ and the prediction $\mathbf P_i\in \mathbb{R}^{1\times H_i \times W_i}$, the extracted lesion representations by Equ.~\ref{equ1} (w/o $up_2$) serve as the $\mathbf K$ and $\mathbf V \in \mathbb{R}^{1 \times C_i}$ to perform the cross-attention, which is similar to the above mentioned self-attention. Through such an operation, the more similar the region to the lesion, the further enhancement of lesion characteristics, which significantly improves the feature contrast and benefits to detect concealed polyps.
\section{Experiments}
\subsection{Datasets}
\textbf{Public Polyp Benchmarks} We evaluate our proposed LDNet on four public polyp datasets, including Kvasir-SEG~\cite{jha2020kvasir}, CVC-ClinicDB~\cite{bernal2015wm}, CVC-ColonDB~\cite{tajbakhsh2015automated} and ETIS~\cite{silva2014toward}. Following the same setting in~\cite{wei2021shallow,fan2020pranet}, we randomly select $80\%$ images respectively from Kvasir-SEG and CVC-ClinicDB and fuse them together as our training set, $10\%$ as validation set. The remaining data of Kvasir-SEG and CVC-ClinicDB, and other two unseen datasets are used for testing.\\
\textbf{Our Collected Large-Scale Polyp Dataset}
We also evaluate LDNet on our collected polyp dataset, which has 5175 images in total. This dataset is randomly split into 60\% for training, 20\% for validation, and the remaining for testing.

\subsection{Implementation Details and Evaluation Metrics}
Our method is implemented based on PyTorch framework~\cite{paszke2019pytorch} and runs on an NVIDIA GeForce RTX 2080 Ti GPU. We simply set $K=1$ in the kernel generation and $n=8$ in the multi-head attention mechanism. The SGD optimizer is utilized to train the model, with batch size of 8,
momentum of 0.9 and weight decay of $10^{-5}$. The initial learning rate is set to 0.001, and adjusted by a poly learning rate policy, which is $lr=lr_{init}\times(1-\frac{epoch}{nEpoch})^{power}$, where $power=0.9$, $nEpoch=80$.
All images are uniformly resized to $256 \times 256$. To avoid overfitting, data augmentations including random horizontal and vertical flips, rotation, random cropping are used in the training stage. A combination of Binary Cross-Entropy loss and Dice loss is used to supervise the training process.

As in~\cite{zhang2020adaptive,fang2019selective}, eight common metrics are adopted to evaluate polyp segmentation performance, including \emph{Recall}, \emph{Specificity}, \emph{Precision}, \emph{Dice Score}, \emph {IoU for Polyp (IoUp)}, \emph{IoU for Background (IoUb)}, \emph{Mean IoU (mIoU)} and \emph {Accuracy}.

\begin{table*}[tp]
\centering
\caption{Comparison with other state-of-the-art methods on four benchmark datasets. The best three results are highlighted in \textcolor{red}{red}, \textcolor{green}{green} and \textcolor{blue}{blue}, respectively.}
\resizebox{\textwidth}{!}{
\begin{tabular}{p{40pt}|p{65pt}|p{30pt}|p{30pt}|p{30pt}|p{30pt}|p{30pt}|p{30pt}|p{30pt}|p{30pt}} 
\hline
& Methods & $Rec$  & $Spec$ & $Prec$ & $Dice$ & $IoUp$ & $IoUb$ & $mIoU$ & $Acc$ \\
\hline
\multirow{8}*{Kvasir}&UNet~\cite{ronneberger2015u}&87.04&97.25&84.28&82.60&73.39&93.89&83.64&95.05 \\
&ResUNet~\cite{zhang2018road} &84.70&97.17&83.00&80.50&70.60&93.19&81.89&94.43 \\
&UNet++~\cite{zhou2018unet++} &89.23&97.20&85.57&84.77&76.42&94.23&85.32&95.44 \\
&ACSNet~\cite{zhang2020adaptive}&91.35&\textcolor{red}{98.39}&\textcolor{green}{91.46}&89.54&83.72&\textcolor{blue}{96.42}&\textcolor{blue}{90.07}&\textcolor{blue}{97.16} \\
&PraNet~\cite{fan2020pranet} &\textcolor{red}{93.90}&97.33&89.87&\textcolor{green}{90.32}&\textcolor{green}{84.55}&95.98&\textcolor{green}{90.26}&96.75 \\
&CCBANet~\cite{nguyen2021ccbanet} &90.71&98.04&91.02&89.04&82.82&96.21&89.52&97.02 \\
&SANet~\cite{wei2021shallow} &\textcolor{blue}{92.06}&\textcolor{green}{98.20}&\textcolor{blue}{91.14}&\textcolor{blue}{89.92}&\textcolor{blue}{83.97}&\textcolor{green}{96.54}&\textcolor{green}{90.26}&\textcolor{green}{97.18} \\
&\textbf{Ours}  &\textcolor{green}{92.72}&\textcolor{blue}{98.05}&\textcolor{red}{92.04}&\textcolor{red}{90.70}&\textcolor{red}{85.30}&\textcolor{red}{96.71}&\textcolor{red}{91.01}&\textcolor{red}{97.35} \\
\hline

\multirow{8}*{\shortstack{CVC-\\ClinicDB}}&UNet~\cite{ronneberger2015u}&88.61&98.70&85.10&85.12&77.78&97.70&87.74&97.95 \\
&ResUNet~\cite{zhang2018road} &90.89&99.25&90.22&89.98&82.77&98.18&90.47&98.37 \\
&UNet++~\cite{zhou2018unet++} &87.78&99.21&90.02&87.99&80.69&97.92&89.30&98.12 \\
&ACSNet~\cite{zhang2020adaptive}&93.46&\textcolor{red}{99.54}&\textcolor{red}{94.63}&\textcolor{green}{93.80}&\textcolor{green}{88.57}&\textcolor{red}{98.95}&\textcolor{green}{93.76}&\textcolor{red}{99.08} \\
&PraNet~\cite{fan2020pranet} &\textcolor{red}{95.22}&99.34&92.25&93.49&88.08&\textcolor{blue}{98.92}&93.50&\textcolor{blue}{99.05} \\
&CCBANet~\cite{nguyen2021ccbanet} &\textcolor{green}{94.89}&99.22&91.39&92.83&86.96&98.79&92.87&98.93 \\
&SANet~\cite{wei2021shallow} &\textcolor{blue}{94.74}&\textcolor{blue}{99.41}&\textcolor{blue}{92.88}&\textcolor{blue}{93.61}&\textcolor{blue}{88.26}&\textcolor{green}{98.94}&\textcolor{blue}{93.60}&\textcolor{green}{99.07} \\
&\textbf{Ours}  &94.49&\textcolor{green}{99.51}&\textcolor{green}{94.53}&\textcolor{red}{94.31}&\textcolor{red}{89.48}&\textcolor{red}{98.95}&\textcolor{red}{94.21}&\textcolor{red}{99.08} \\
\hline

\multirow{8}*{\shortstack{CVC-\\ColonDB}}&UNet~\cite{ronneberger2015u}&63.05&98.00&68.01&56.40&47.32&94.51&70.92&94.84 \\
&ResUNet~\cite{zhang2018road} &59.91&98.06&65.29&54.87&44.31&93.77&69.04&94.06 \\
&UNet++~\cite{zhou2018unet++} &63.49&\textcolor{blue}{98.59}&77.79&60.77&52.64&95.19&73.92&95.48 \\
&ACSNet~\cite{zhang2020adaptive}&77.38&\textcolor{red}{99.26}&\textcolor{red}{81.72}&\textcolor{blue}{75.51}&\textcolor{blue}{67.38}&\textcolor{blue}{96.16}&\textcolor{blue}{81.77}&\textcolor{blue}{96.32} \\
&PraNet~\cite{fan2020pranet} &\textcolor{blue}{81.85}&98.54&78.43&\textcolor{green}{76.24}&\textcolor{green}{68.29}&96.06&\textcolor{green}{82.17}&96.26 \\
&CCBANet~\cite{nguyen2021ccbanet} &\textcolor{green}{82.34}&98.39&77.79&75.36&66.57&95.89&81.23&96.14 \\
&SANet~\cite{wei2021shallow} &75.21&\textcolor{green}{99.09}&\textcolor{green}{81.43}&73.50&65.47&\textcolor{green}{96.19}&80.83&\textcolor{green}{96.40} \\
&\textbf{Ours}  &\textcolor{red}{83.46}&98.49&\textcolor{blue}{81.15}&\textcolor{red}{78.43}&\textcolor{red}{70.58}&\textcolor{red}{96.21}&\textcolor{red}{83.39}&\textcolor{red}{96.48} \\
\hline

\multirow{8}*{ETIS}&UNet~\cite{ronneberger2015u}&47.33&96.36&48.05&34.81&28.38&94.72&61.55&94.90 \\
&ResUNet~\cite{zhang2018road} &49.12&97.21&56.85&38.65&30.54&95.27&62.90&95.43 \\
&UNet++~\cite{zhou2018unet++} &55.52&95.40&59.14&40.91&33.86&93.87&63.87&94.07 \\
&ACSNet~\cite{zhang2020adaptive}&78.31&\textcolor{blue}{98.44}&68.81&69.44&60.96&97.78&79.37&97.89 \\
&PraNet~\cite{fan2020pranet} &\textcolor{green}{81.20}&\textcolor{green}{98.73}&\textcolor{green}{72.23}&\textcolor{green}{72.38}&\textcolor{green}{64.07}&\textcolor{green}{98.29}&\textcolor{blue}{81.18}&\textcolor{green}{98.38} \\
&CCBANet~\cite{nguyen2021ccbanet} &\textcolor{blue}{78.70}&97.19&61.12&62.63&53.81&96.52&75.17&96.66 \\
&SANet~\cite{wei2021shallow} &77.08&\textcolor{red}{99.04}&\textcolor{red}{72.73}&\textcolor{blue}{72.26}&\textcolor{blue}{63.33}&\textcolor{red}{98.47}&\textcolor{green}{80.90}&\textcolor{red}{98.54} \\
&\textbf{Ours}  &\textcolor{red}{82.83}&\textcolor{blue}{98.44}&\textcolor{blue}{72.07}&\textcolor{red}{74.37}&\textcolor{red}{66.50}&\textcolor{blue}{98.01}&\textcolor{red}{82.26}&\textcolor{blue}{98.10} \\
\hline
\end{tabular}}
\label{table1}
\end{table*}

\begin{table*}
\centering
\caption{Comparison with other state-of-the-art methods and ablation study on our collected dataset.}
\resizebox{\textwidth}{!}{
\begin{tabular}{p{100pt}|p{30pt}|p{30pt}|p{30pt}|p{30pt}|p{30pt}|p{30pt}|p{30pt}|p{30pt}} 
\toprule
Methods & $Rec$  & $Spec$ & $Prec$ & $Dice$ & $IoUp$ & $IoUb$ & $mIoU$ & $Acc$ \\
\midrule
UNet~\cite{ronneberger2015u}&87.89&97.27&87.23&85.00&77.48&93.95&85.71&95.64 \\
UNet++~\cite{zhou2018unet++} &89.88&97.43&88.18&86.92&79.88&94.56&87.26&96.21 \\
ACSNet~\cite{zhang2020adaptive} &\textcolor{blue}{92.43}&97.79&90.94&90.54&84.64&95.75&90.19&97.11 \\
PraNet~\cite{fan2020pranet} &\textcolor{green}{92.86}&97.87&90.52&90.64&84.60&\textcolor{blue}{95.91}&90.25&\textcolor{green}{97.28} \\
CCBANet~\cite{nguyen2021ccbanet}&91.91&97.79&91.32&90.39&84.36&95.73&90.04&97.10 \\
SANet~\cite{wei2021shallow} &92.18&\textcolor{green}{98.22}&\textcolor{blue}{91.67}&\textcolor{green}{90.75}&\textcolor{green}{84.98}&\textcolor{green}{96.02}&\textcolor{green}{90.50}&\textcolor{blue}{97.27} \\
\hline
\textbf{Ours}  &\textcolor{red}{93.22}&\textcolor{blue}{98.15}&\textcolor{red}{92.16}&\textcolor{red}{91.66}&\textcolor{red}{86.28}&\textcolor{red}{96.39}&\textcolor{red}{91.34}&\textcolor{red}{97.55} \\
\hline
Baseline&92.02&97.03&87.75&88.30&81.26&94.95&88.11&96.54\\
Baseline+DK &92.22&97.58&90.41&89.88&83.86&95.53&89.70&96.92\\
Baseline+DK+ESAs&91.76&\textcolor{red}{98.25}&\textcolor{green}{92.14}&\textcolor{blue}{90.74}&\textcolor{blue}{84.91}&95.85&\textcolor{blue}{90.38}&97.14\\
\bottomrule
\end{tabular}}
\label{table2}
\end{table*}

\subsection{Experiments on the Public Polyp Benchmarks}
\begin{figure}[tp]
\centering
\includegraphics[width=0.95\textwidth]{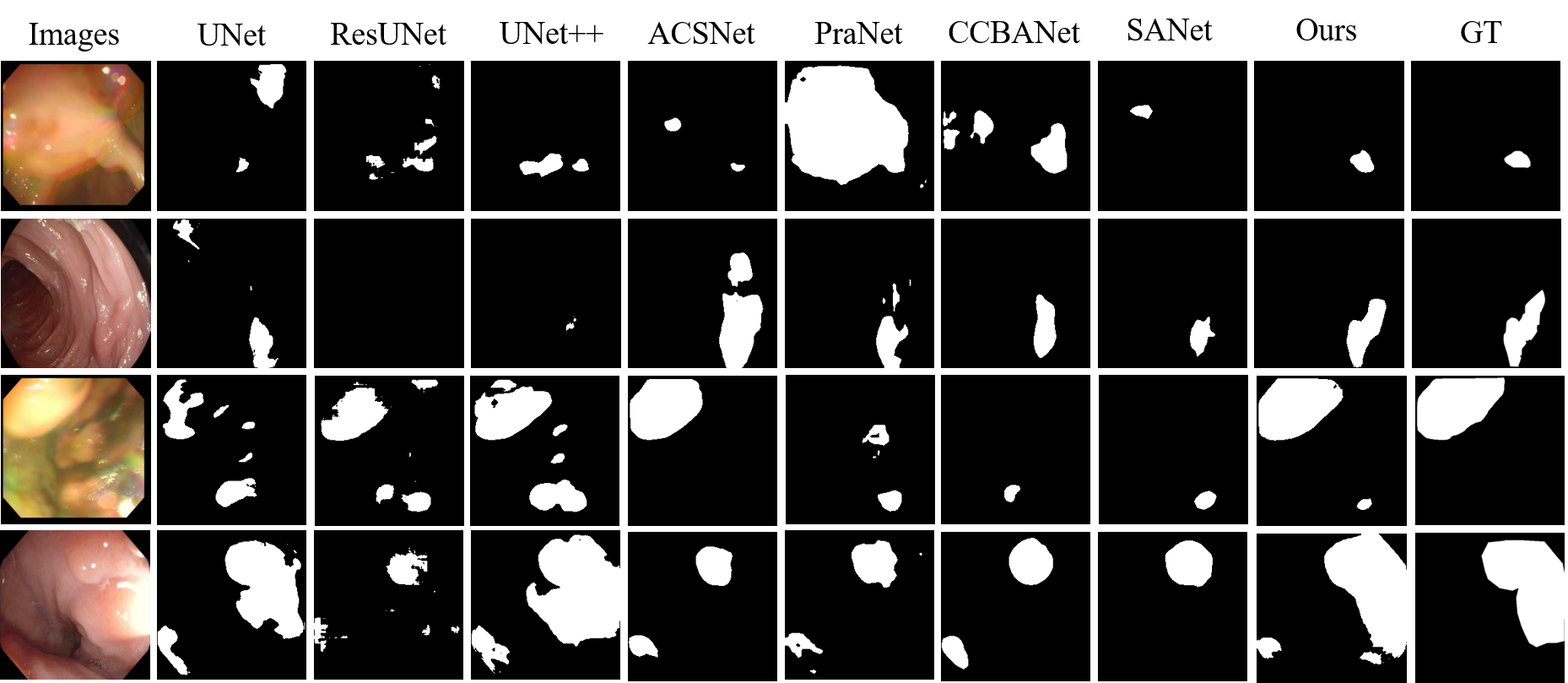} 
\caption{Visual comparison of polyp segmentation results.}
\label{fig3}
\end{figure}

We compare our LDNet with several state-of-the-art methods, including UNet~\cite{ronneberger2015u}, ResUNet~\cite{zhang2018road}, UNet++\cite{zhou2018unet++}, ACSNet~\cite{zhang2020adaptive}, PraNet~\cite{fan2020pranet}, SANet~\cite{wei2021shallow}, CCBANet~\cite{nguyen2021ccbanet}, on the public polyp benchmarks. As shown in Table~\ref{table1}, our LDNet achieves superior performance over other methods across four datasets on most metrics. In particular, on the two seen datasets, \emph{i.e.,} Kvasir and CVC-ClinicDB, the proposed LDNet obtains the best \emph{Dice} and \emph{mIoU} scores, outperforming other methods. On the other two unseen datasets, the LDNet also shows strong generalization ability and achieves $78.43\%$ and $74.37\%$ \emph{Dice} scores, $2.19\%$ and $1.99\%$ improvements over the second best approaches, further demonstrating the effectiveness of our approach. Some visualization examples are shown in Fig.~\ref{fig3}.

\subsection{Experiments on the Collected Large-Scale Polyp Dataset}
On our collected large-scale polyp dataset, we compare the LDNet with UNet~\cite{ronneberger2015u}, UNet++\cite{zhou2018unet++}, ACSNet~\cite{zhang2020adaptive}, PraNet~\cite{fan2020pranet}, SANet~\cite{wei2021shallow} and CCBANet~\cite{nguyen2021ccbanet}. As shown in Table~\ref{table2}, our method again achieves the best performance, with a \emph{Dice} of $91.66\%$ and a \emph{mIoU} of $91.34\%$, respectively.

\subsection{Ablation Study}
We conduct a series of ablation studies on our collected polyp dataset to verify the effectiveness of our designed dynamic kernel scheme and attention modules. Specifically, we utilize the traditional u-shape structure with a static segmentation head as our baseline, and gradually replace the static head with our designed dynamic kernels, then further add ESA and LCA modules, denoting as Baseline, Baseline+DK, Baseline+DK+ESAs and Ours respectively. As shown in Table~\ref{table2}, the introduction of the dynamic kernel significantly 
enhances the performance of the baseline, with a $1.58\%$ improvement of \emph{Dice} score. With the addition of our ESA and LCA modules, the scores of \emph{Dice} and \emph{mIoU} are further boosted by $0.86\%$ and $0.68\%$, $0.92\%$ and $0.96\%$, respectively.
\section{Conclusion}
In this paper, we propose the lesion-aware dynamic kernel (LDNet) for polyp segmentation, which is generated conditioned on the global information and updated by the multi-level lesion features. We believe that such a dynamic kernel can endow our model with more flexibility to attend diverse polyps regions. Besides, we also improve the feature representation and enhance the context contrast by two tailored attention modules, \emph{i.e.,} ESA and LCA, which is beneficial for detecting concealed polyps. Extensive experiments and ablation studies demonstrate the effectiveness of our proposed method.

\subsubsection{Acknowledgements}
This work is supported in part by the Chinese Key-Area Research and Development Program of Guangdong Province~(2020B0101350001), in part by the Guangdong Basic and Applied Basic Research Foundation~(2020B\\1515020048), in part by the National Natural Science Foundation of China (61976250), in part by the Guangzhou Science and technology project (20210202\\0633), and is also supported by the Guangdong Provincial Key Laboratory of Big Data Computing, The Chinese University of Hong Kong, Shenzhen.

%
%
%
\bibliographystyle{splncs04}
\bibliography{paper847}

\begin{thebibliography}{10}
\providecommand{\url}[1]{\texttt{#1}}
\providecommand{\urlprefix}{URL }
\providecommand{\doi}[1]{https://doi.org/#1}

\bibitem{akbari2018polyp}
Akbari, M., et~al.: Polyp segmentation in colonoscopy images using fully
  convolutional network. In: 2018 40th Annual International Conference of the
  IEEE Engineering in Medicine and Biology Society. pp. 69--72 (2018)

\bibitem{bernal2015wm}
Bernal, J., S{\'a}nchez, F.J., Fern{\'a}ndez-Esparrach, G., Gil, D.,
  Rodr{\'\i}guez, C., Vilari{\~n}o, F.: Wm-dova maps for accurate polyp
  highlighting in colonoscopy: Validation vs. saliency maps from physicians.
  Computerized Medical Imaging and Graphics  \textbf{43},  99--111 (2015)

\bibitem{brandao2017fully}
Brandao, P., et~al.: Fully convolutional neural networks for polyp segmentation
  in colonoscopy. In: Medical Imaging 2017: Computer-Aided Diagnosis. vol.
  10134, p. 101340F. International Society for Optics and Photonics (2017)

\bibitem{carion2020end}
Carion, N., Massa, F., Synnaeve, G., Usunier, N., Kirillov, A., Zagoruyko, S.:
  End-to-end object detection with transformers. In: Vedaldi, A., et~al. (eds.)
  ECCV 2020. LNCS, vol. 12346, pp. 213--229. Springer, Cham (2020).
  \doi{10.1007/978-3-030-58452-8\_13}

\bibitem{dosovitskiy2020image}
Dosovitskiy, A., Beyer, L., Kolesnikov, A., Weissenborn, D., Zhai, X.,
  Unterthiner, T., Dehghani, M., Minderer, M., Heigold, G., Gelly, S., et~al.:
  An image is worth 16x16 words: Transformers for image recognition at scale.
  arXiv preprint arXiv:2010.11929  (2020)

\bibitem{fan2020pranet}
Fan, D.P., Ji, G.P., Zhou, T., Chen, G., Fu, H., Shen, J., Shao, L.: Pranet:
  Parallel reverse attention network for polyp segmentation. In: Martel, A.L.,
  et~al. (eds.) MICCAI 2020. LNCS, vol. 12266, pp. 263--273. Springer, Cham
  (2020). \doi{10.1007/978-3-030-59725-2\_26}

\bibitem{fang2019selective}
Fang, Y., Chen, C., Yuan, Y., Tong, K.y.: Selective feature aggregation network
  with area-boundary constraints for polyp segmentation. In: Shen, D., et~al.
  (eds.) MICCAI 2019. LNCS, vol. 11764, pp. 302--310. Springer, Cham (2019).
  \doi{10.1007/978-3-030-32239-7\_34}

\bibitem{gao2019res2net}
Gao, S.H., Cheng, M.M., Zhao, K., Zhang, X.Y., Yang, M.H., Torr, P.: Res2net: A
  new multi-scale backbone architecture. IEEE transactions on pattern analysis
  and machine intelligence  \textbf{43}(2),  652--662 (2019)

\bibitem{he2019dynamic}
He, J., Deng, Z., Qiao, Y.: Dynamic multi-scale filters for semantic
  segmentation. In: Proceedings of the IEEE/CVF International Conference on
  Computer Vision. pp. 3562--3572 (2019)

\bibitem{jha2020kvasir}
Jha, D., et~al.: Kvasir-seg: A segmented polyp dataset. In: Ro, Y.M., et~al.
  (eds.) MMM 2020. LNCS, vol. 11962, pp. 451--462. Springer, Cham (2020).
  \doi{10.1007/978-3-030-37734-2\_37}

\bibitem{long2015fully}
Long, J., Shelhamer, E., Darrell, T.: Fully convolutional networks for semantic
  segmentation. In: Proceedings of the IEEE Conference on Computer Vision and
  Pattern Recognition. pp. 3431--3440 (2015)

\bibitem{mamonov2014automated}
Mamonov, A.V., Figueiredo, I.N., Figueiredo, P.N., Tsai, Y.H.R.: Automated
  polyp detection in colon capsule endoscopy. IEEE transactions on medical
  imaging  \textbf{33}(7),  1488--1502 (2014)

\bibitem{nguyen2021ccbanet}
Nguyen, T.C., Nguyen, T.P., Diep, G.H., Tran-Dinh, A.H., Nguyen, T.V., Tran,
  M.T.: Ccbanet: Cascading context and balancing attention for polyp
  segmentation. In: de~Bruijne, M., et~al. (eds.) MICCAI 2021. LNCS, vol.
  12901, pp. 633--643. Springer, Cham (2021).
  \doi{10.1007/978-3-030-87193-2\_60}

\bibitem{pang2020hierarchical}
Pang, Y., Zhang, L., Zhao, X., Lu, H.: Hierarchical dynamic filtering network
  for rgb-d salient object detection. In: Vedaldi, A., et~al. (eds.) ECCV 2020.
  LNCS, vol. 12370, pp. 235--252. Springer, Cham (2020).
  \doi{10.1007/978-3-030-58595-2\_15}

\bibitem{paszke2019pytorch}
Paszke, A., et~al.: Pytorch: An imperative style, high-performance deep
  learning library. In: Advances in Neural Information Processing Systems. pp.
  8026--8037 (2019)

\bibitem{ronneberger2015u}
Ronneberger, O., Fischer, P., Brox, T.: U-net: Convolutional networks for
  biomedical image segmentation. In: Navab, N., Hornegger, J., Wells, W.M.,
  Frangi, A.F. (eds.) MICCAI 2015. LNCS, vol.~9351, pp. 234--241. Springer,
  Cham (2015). \doi{10.1007/978-3-319-24574-4\_28}

\bibitem{Siegel2022cancer}
Siegel, R.L., Miller, K.D., Fuchs, H.E., Jemal, A.: Cancer statistics, 2022.
  CA: A Cancer Journal for Clinicians  \textbf{72}(1),  7--33 (2022)

\bibitem{silva2014toward}
Silva, J., Histace, A., Romain, O., Dray, X., Granado, B.: Toward embedded
  detection of polyps in wce images for early diagnosis of colorectal cancer.
  International journal of computer assisted radiology and surgery
  \textbf{9}(2),  283--293 (2014)

\bibitem{tajbakhsh2015automated}
Tajbakhsh, N., Gurudu, S.R., Liang, J.: Automated polyp detection in
  colonoscopy videos using shape and context information. IEEE transactions on
  medical imaging  \textbf{35}(2),  630--644 (2015)

\bibitem{vaswani2017attention}
Vaswani, A., Shazeer, N., Parmar, N., Uszkoreit, J., Jones, L., Gomez, A.N.,
  Kaiser, {\L}., Polosukhin, I.: Attention is all you need. Advances in neural
  information processing systems  \textbf{30} (2017)

\bibitem{wei2021shallow}
Wei, J., Hu, Y., Zhang, R., Li, Z., Zhou, S.K., Cui, S.: Shallow attention
  network for polyp segmentation. In: de~Bruijne, M., et~al. (eds.) MICCAI
  2021. LNCS, vol. 12901, pp. 699--708. Springer, Cham (2021).
  \doi{10.1007/978-3-030-87193-2\_66}

\bibitem{zhang2021dodnet}
Zhang, J., Xie, Y., Xia, Y., Shen, C.: Dodnet: Learning to segment multi-organ
  and tumors from multiple partially labeled datasets. In: Proceedings of the
  IEEE/CVF Conference on Computer Vision and Pattern Recognition. pp.
  1195--1204 (2021)

\bibitem{zhang2020adaptive}
Zhang, R., Li, G., Li, Z., Cui, S., Qian, D., Yu, Y.: Adaptive context
  selection for polyp segmentation. In: Martel, A.L., et~al. (eds.) MICCAI
  2020. LNCS, vol. 12266, pp. 253--262. Springer, Cham (2020).
  \doi{10.1007/978-3-030-59725-2\_25}

\bibitem{zhang2021k}
Zhang, W., Pang, J., Chen, K., Loy, C.C.: K-net: Towards unified image
  segmentation. Advances in Neural Information Processing Systems  \textbf{34}
  (2021)

\bibitem{zhang2018road}
Zhang, Z., Liu, Q., Wang, Y.: Road extraction by deep residual u-net. IEEE
  Geoscience and Remote Sensing Letters  \textbf{15}(5),  749--753 (2018)

\bibitem{zhao2017pyramid}
Zhao, H., Shi, J., Qi, X., Wang, X., Jia, J.: Pyramid scene parsing network.
  In: Proceedings of the IEEE Conference on Computer Vision and Pattern
  Recognition. pp. 2881--2890 (2017)

\bibitem{zhou2018unet++}
Zhou, Z., Siddiquee, M.M.R., Tajbakhsh, N., Liang, J.: Unet++: A nested u-net
  architecture for medical image segmentation. In: Stoyanov, D., et~al. (eds.)
  DLMIA/ML-CDS 2018. LNCS, vol. 11045, pp. 3--11. Springer, Cham (2018).
  \doi{10.1007/978-3-030-00889-5\_1}

\bibitem{zhu2019asymmetric}
Zhu, Z., Xu, M., Bai, S., Huang, T., Bai, X.: Asymmetric non-local neural
  networks for semantic segmentation. In: Proceedings of the IEEE/CVF
  International Conference on Computer Vision. pp. 593--602 (2019)

\end{thebibliography}

\end{document}